\documentclass{sigplanconf}

\usepackage{hyperref}
\usepackage{doi}
\usepackage{amsmath,amsthm,amssymb}
\usepackage{bm}
\usepackage{hypernat}
\usepackage{xcolor}
\usepackage[utf8]{inputenc}
\usepackage{listings}
\usepackage{booktabs}
\usepackage{placeins}

\usepackage{tikz}
\usepackage{tikzscale}
\usepackage[skins,listings]{tcolorbox}

\usepackage[binary-units]{siunitx}
\DeclareSIUnit{\flops}{ FLOP\per\second }

\tcbset{listing engine=listings}
\tcbset{listingbox/.style={%
    enhanced,
    boxsep=-9pt,
    left=15pt,
    arc=2pt,
    boxrule=1pt,
  }
}

\definecolor{green}{RGB}{0, 180, 0}

\lstdefinestyle{custompython}{%
    belowcaptionskip=1\baselineskip,
    breaklines=true,
    frame=none,
    xleftmargin=0pt,
    language=Python,
    showstringspaces=false,
    basicstyle=\footnotesize\ttfamily,
    keywordstyle=\bfseries\color{green!40!black},
    commentstyle=\itshape\color{purple},
    identifierstyle=\color{black},
    stringstyle=\color{green!60!black},
    keywords=[2]{precompute,assignment_to_subst,split_iname,
      tag_data_axes,split_array_dim,tag_data_axes,set_array_dim_name,
      set_array_dim_names,tag_inames,fix_parameters,fuse_kernels,assume,
      set_loop_priority,alias_temporaries,rename_iname,
      collect_common_factors_on_increment,buffer_array,add_prefetch},
    keywordstyle=[2]\bfseries\color{blue},
    numbers=none,
    rangebeginprefix=\#,
    rangeendprefix=\#,
    columns=fullflexible,
    includerangemarker=false
}
\lstdefinestyle{customxform}{%
    belowcaptionskip=1\baselineskip,
    breaklines=true,
    frame=none,
    xleftmargin=0pt,
    language=Fortran,
    showstringspaces=false,
    basicstyle=\footnotesize\ttfamily,
    keywordstyle=\bfseries\color{green!40!black},
    commentstyle=\itshape\color{purple},
    identifierstyle=\color{black},
    stringstyle=\color{green!60!black},
    numbers=none,
    columns=flexible,
    rangebeginprefix=!\ \#,
    rangeendprefix=!\ \#,
    includerangemarker=false
}
\lstdefinestyle{customfortran}{%
    belowcaptionskip=1\baselineskip,
    breaklines=true,
    frame=none,
    xleftmargin=0pt,
    language=Fortran,
    showstringspaces=false,
    basicstyle=\footnotesize\ttfamily,
    keywordstyle=\bfseries\color{green!60!black},
    commentstyle=\itshape\color{purple},
    identifierstyle=\color{black},
    stringstyle=\color{green},
    numbers=none,
    columns=flexible,
    rangebeginprefix=!\ ,
    rangeendprefix=!\ ,
    includerangemarker=false
}
\lstdefinestyle{customc}{%
    belowcaptionskip=1\baselineskip,
    breaklines=true,
    frame=none,
    xleftmargin=\parindent,
    language=C,
    showstringspaces=false,
    basicstyle=\footnotesize\ttfamily,
    keywordstyle=\bfseries\color{green!60!black},
    commentstyle=\itshape\color{purple},
    identifierstyle=\color{black},
    stringstyle=\color{green},
    numbers=none,
    columns=flexible,
}

\def\loopy{Loo.py}

\newcommand{\trans} {\ensuremath{\intercal}}
\renewcommand{\vec}[1] {\ensuremath{\bm{#1}}%
}

\newcommand{\pder} [2]{\frac{\partial#1}{\partial#2}}

\newcommand{\sd}[1]{\ensuremath{\mathrm{#1}}%
}

\newcommand\prexformincl[1]{%
\lstinputlisting[style=customxform,linerange=START_XFORM_#1-END_XFORM_#1]{strongVolumeKernels.f90}%
}
\newcommand\xformincl[1]{%
\lstinputlisting[style=custompython,linerange=START_XFORM_#1-END_XFORM_#1]{strongVolumeKernels-transform.py}%
}

\def\optlevel#1{\noindent[$\nwarrow$ Opt. level #1]}

\begin{document}


\setlength{\pdfpageheight}{\paperheight}
\setlength{\pdfpagewidth}{\paperwidth}

\conferenceinfo{ARRAY'16}{June 14, 2016, Santa Barbara, CA, USA}
\CopyrightYear{2014}
\crdata{}
\sigdoi{}


\exclusivelicense%


\titlebanner{Program Transformation \loopy}        
\preprintfooter{Submitted to \href{http://www.sable.mcgill.ca/array/}{ARRAY16}}   

\title{%
Array Program Transformation with \loopy\ by Example: High-Order Finite Elements}

\authorinfo{Andreas Klöckner}
           {University of Illinois at Urbana-Champaign, Urbana, IL, USA}
           {andreask@illinois.edu}
\authorinfo{Lucas~C. Wilcox}
           {Naval Postgraduate School, Monterey, CA, USA}
           {lwilcox@nps.edu}
\authorinfo{T. Warburton}
           {Virginia Polytechnic Institute and State University, Blacksburg, VA, USA}
           {tim.warburton@vt.edu}

\maketitle

\begin{abstract}
  To concisely and effectively demonstrate the capabilities of our program
  transformation system  \loopy, we examine a transformation path from two
  real-world Fortran subroutines as found in a weather model to a single
  high-performance computational kernel suitable for execution on modern GPU
  hardware. Along the transformation path, we encounter kernel fusion,
  vectorization, prefetching, parallelization, and algorithmic changes achieved
  by mechanized conversion between imperative and functional/substitution-based
  code, among a number more.
  We conclude with performance results that demonstrate the
  effects and support the effectiveness of the applied transformations.
\end{abstract}

\category{D.3.4}{Programming Languages}{Processors}  --- Code generators
\category{D.1.3}{Programming Languages}{Programming Techniques} --- Concurrent programming
\category{G.4}{Mathematics of Computing}{Mathematical software}


\keywords%
Code generation, high-level language, GPU, substitution rule,
embedded language, high-performance, program transformation,
OpenCL


\section{Introduction}
\label{sec:intro}
User-guided transformation of numerical and array-based computations is an area
of sustained interest across the areas of high-performance computing,
programming languages, and numerical methods. Its existence is inspired by the
discrepancy between a compiler's hypothetical ability to transform programs in
ways that would ostensibly be beneficial to performance and their practical
inability to do so as hampered by (1) the size of the search space of such
transformations, and (2) a compiler's ability to prove such transformations
correct and/or equivalent to prior behavior.

Descriptions of such systems in the literature often restrict themselves to
simple, common, ``micro-benchmark'' examples. It is understandable that this
approach is seeing much use, since it is the least demanding in terms of the
reader's attention span, the required space, and, also, the implemented
capability of the transformation system. Further, it facilitates easy
comparison of performance results between systems and allows for the presentation
of many examples. This article is an experiment in the opposite
approach. We present a single, realistic example of program transformation
drawn from the application domain of weather prediction. In taking this
approach, we hope to better showcase capabilities present in our
transformation system and the language mechanisms present that enable
their use, all while describing the trade-offs that lead to the actual
implemented design.  We hope that others may follow our example and
that a body of literature may arise that can serve to motivate and
guide the discussion on program transformation.

\loopy\ \citep{kloeckner_loopy_2014,kloeckner_loopy_2015} is a programming system for array
computations that targets CPUs, GPUs, and other, potentially heterogeneous
compute architectures. One salient feature of \loopy\ is that programs
written in it necessarily consist of two parts:
\begin{itemize}
\item A semi-mathematical statement of the array computation to be carried
  out, in terms of a \emph{loop polyhedron} and a partially ordered \emph{set
  of `instructions'}.
\item A sequence of \emph{kernel transformations}, driven by an `outer'
  program in the high-level scripting language Python \citep{vanrossum_python_1994}.
\end{itemize}
This strong separation is an explicit design goal, as it enables specialization of
users, cleanliness of notation in either part, as well as greater
flexibility in terms of transformation.

The present article demonstrates how \loopy\ can function as a code generation
and transformation engine for computational code originally expressed in a
subset of Fortran while maintaining its full capability to transform the
ingested code in a manner comprehensible and useful to the author of the
original program. A number of mechanisms are described that are intended to aid
the formulation of transformations on array computations in this setting.
As one example of the issues that arise, the strong separation of semantics
and transformation, while desirable, also poses a difficulty. For example,
unlike in an annotation-based setting, where lexical proximity alone can be
used to indicate what part of a program is to be transformed, this option
does not exist for \loopy, and so alternatives have to be devised.

\subsection{Related work on code generation}
The literature on code generation and optimization for array
languages is vast, and no attempt will be made to provide a survey of the
subject in any meaningful way. Instead, we will seek to highlight a few
approaches that have significantly influenced the thinking behind \loopy,
are particularly similar, or provide ideas for further development.
\loopy\ is heavily inspired by the polyhedral model of expressing
static-control programs~\citep{feautrier_automatic_1996,bastoul_code_2004}.
While it takes significant inspiration from this approach, the details of
how a program is represented, beyond the existence of a loop domain, are
quite different.  High-performance compilation for GPUs, by now, is hardly
a new topic, and many different approaches have been used, including
ones using OpenMP-style
directives~\citep{lee_openmpc_2010,han_hicuda_2011},
ones that are fully automatic~\citep{yang_gpgpu_2010},
ones based on functional languages~\citep{svensson_obsidian_2010}, and
ones based on the polyhedral model~\citep{verdoolaege_polyhedral_2013}.
Other ones
define an automatic, array computation
middleware~\citep{garg_velociraptor_2012} designed as a back-end for
multiple languages,
including Python. Automatic, GPU-targeted compilers for languages embedded
in Python also
abound~\citep{catanzaro_copperhead_2011,rubinsteyn_parakeet_2012,continuum_numba_2014},
most of which transform a Python AST at run-time based on various
levels of annotation and operational abstraction.

User-guided program transformation based on polyhedral representation has
received considerable attention over the years.
Perhaps the conceptually closest prior work to the approach taken by
\loopy\ is CUDA-CHiLL~\citep{rudy_programming_2011}, which performs
source-to-source translation based on a set of user-controlled
transformations~\citep{chen_chill_2008,hall_loop_2010}. \loopy\ and CHiLL
still are not quite alike, using dissimilar intermediate representations,
dissimilar levels of abstraction in the description of transformations, and
a dissimilar (static vs.\ program-controlled) approach to transformation.
Other similar projects include the AlphaZ
\citep{yuki_alphaz_2012} and Clay systems \citep{bagneres_opening_2016},
although these projects emphasize the scheduling of a given workload rather than its
algorithmic or data-based transformation. Rewriting- and substitution-focused
systems such as Terra~\citep{devito_terra_2013}, like \loopy, provide a powerful
building block for DSLs, but they lack the loop transformation and parallelization
capabilities afforded by polyhedral representation.

Other optimizing compilers assume a substantial amount of domain knowledge
(such as what is needed for assembly of finite element matrices) and
leverage this to obtain parallel, optimized code. One example of this
family of code generators is COFFEE~\cite{luporini_crossloop_2015}.

Source-to-source transformation similarly has been studied extensively,
with many mature systems existing in the literature
(see for instance~\citet{schordan_source_2003} and
\citet{dave_cetus_2009}).

\subsection{Related work on our example benchmark}
Our benchmark detailed below is an embodiment of the `unstructured grid' `dwarf' of
Colella's oft-cited seven dwarfs~\citep{Colella2004}.  Specifically, we
consider a subclass of finite element models, the
continuous~\citep{Patera1984} and discontinuous spectral element
methods~\citep{Black1999}, which are well suited for simulating wave
phenomena such as acoustics, elastodynamics,
electromagnetics, and fluid dynamics in complex geometry.
We focus on a computation from NUMA (Nonhydrostatic
Unified Model of the Atmosphere)~\citep{KellyGiraldo2012}, the
dynamical core of the U.S. Navy's next generation nonhydrostatic
atmospheric prediction system, NEPTUNE (Navy Environmental Prediction
SysTem Utilizing the NUMA
corE)~\citep{DoyleFY15,GabersekFY14,GabersekFY15}.  Although we focus
on this particular application, the computation would be similar in
other application domains.

The continuous and discontinuous spectral element method has received
attention from the high-performance computing community.  For example,
it has been used by \citet{Tufo1999} and \citet{Komatitsch2003} to win
the ACM/IEEE Supercomputing Gordon Bell prize.  Furthermore, it also
has been ported to GPUs for seismic wave propagation using handwritten
CUDA kernels~\citep{Komatitsch2010,Goeddeke2014,Burstedde2010};
for fluid flow using handwritten OpenCL~\citep{Stilwell2013} and
OpenACC~\citep{Gong2015,Markidis2015} kernels, and using an in-built
domain specific language targeting multiple
backends~\citep{Witherden2015};
for electromagnetic wave propagation using OpenACC~\citep{Otten2016}; and
for elliptic problems using handwritten OCCA~\citep{Medina2014} kernels
targeting multiple backends~\citep{Remacle2015}.
While porting part of the atmospheric climate model CAM-SE from
CUDA Fortran to OpenACC its authors note that ``it is highly unlikely
that a literal single source code would suffice for performance
portability''~\citep{Norman2015}.  This provides motivation to use
programmatic tools, like \loopy, to generate the desired set of
kernels needed for performance potability.

\section{Structure of the Computation}
\label{sec:comp-overview}
To illustrate some of the transformations in \loopy, we consider a
computation performed in a simplified dynamical core of a
nonhydrostatic numerical weather prediction model.  This model is a
discretization of Euler's equations in a method-of-lines approach
using a continuous or discontinuous Galerkin spectral element
discretization in space on a curvilinear hexahedral mesh of
tensor-product polynomial elements (similar to
\citet{KellyGiraldo2012}).  Focusing on explicit time integration, for
our benchmark computation we consider the element volume contribution
to the rate function used by the ordinary differential equation
solver.  This computation is common to both the continuous and
discontinuous Galerkin spectral element methods, which differ by how
the elements are connected.  For a more detailed description of these
methods than provided below, see for example \citet{Kopriva2009}.

For our computation, we consider the governing equations of a dry atmosphere
(without gravity and viscous terms) which are
\begin{equation}\label{set2c}
  \pder{q_{b}}{t} + \sum_{a}\pder{f_{ab}(\vec{q})}{x_{a}} = 0 \quad
\end{equation}
using density, momentum (the three components), potential
temperature density and three tracer densities as prognostic variables
\[\vec{q} = \begin{bmatrix} \rho & U_{1} & U_{2} & U_{3} & \Theta &
                            Q_{1} & Q_{2} & Q_{3} \end{bmatrix}^\trans,\]
where the $a$th column of the flux is
\begin{equation*}
  f_{a}(\vec{q}) = \\ \begin{bmatrix} U_{a} \\
 \frac{U_{a}U_{1}}{\rho} + \delta_{a1} p  \\
 \frac{U_{a}U_{2}}{\rho} + \delta_{a2} p  \\
 \frac{U_{a}U_{3}}{\rho} + \delta_{a3} p  \\
 \frac{U_{a}\Theta}{\rho} \\
 \frac{U_{a}Q_{1}}{\rho} \\
 \frac{U_{a}Q_{2}}{\rho} \\
 \frac{U_{a}Q_{3}}{\rho}
 \end{bmatrix},
\end{equation*}
which is equation set~2C (without gravity and viscous terms) from
\citet{GiraldoRestelliLauter2010}.  Here,
the spatial coordinates are \(\begin{bmatrix} x_{1} & x_{2} &
x_{3} \end{bmatrix}^\trans\), \(t\) is time, \(\delta_{ab}\) is the
Kronecker delta, and \(p\) is the pressure obtained from the equation
of state
\(
  p = p_0 {(\frac{R \Theta}{p_0})}^\gamma,
\)
where \(p_0\) is a constant reference pressure at the surface, \(R=c_p - c_v\)
is the gas constant given in terms of the specific heats for constant pressure
and volume, \(c_p\) and \(c_v\) respectively, and \(\gamma=\frac{c_p}{c_v}\) is
the specific heat ratio.

\begin{figure}
  \centering
  \includegraphics[width=0.5\linewidth]{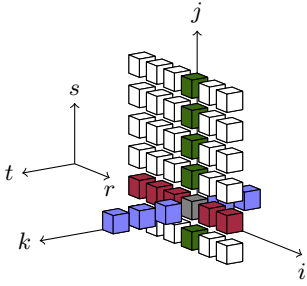}
  \caption{Computational layout of the volume term~\eqref{volterm}.
  Non-white boxes represent the
intra-element grid points involved in the volume term~\eqref{volterm}
associated with the center box; the first term on the right-hand side
involves the (red) boxes in a line parallel to the \(i\)-axis, the
second term involves the (green) boxes in a line parallel to the
\(j\)-axis, and the third term involves (blue) boxes in a line
parallel to the \(k\)-axis.  The group of boxes with the same value of
$k$ is called the $k$-slice.  In addition, element reference
directions $r$, $s$, and $t$ respectively correspond to the loop indices
$i$, $j$, and $k$.}\label{fig:stencil}
\end{figure}

The semi-discretization of Euler's equations~\eqref{set2c} by the
continuous and discontinuous Galerkin methods can be split into
volume, \(\vec{\sd{v}}\), and surface, \(\vec{\sd{s}}\), contributions as
\[
  \pder{\sd{q}_{b}^{eijk}}{t} = \sd{v}_{b}^{eijk}(\vec{\sd{q}})
      + \sd{s}_{b}^{eijk}(\vec{\sd{q}}),
\]
where \(e\) indexes the $N_{e}$ elements and \((i, j, k)\) indexes the
element's $N_{q}^{3}$ grid points.  The surface term connects the
elements together using a `numerical flux' in the discontinuous Galerkin
method and the direct stiffness-summation operator in the continuous
Galerkin method.  As the benchmark in this paper, we consider the
computation of the volume term
\begin{equation}\label{volterm}
\begin{split}
  \sd{v}_{b}^{eijk}(\vec{\sd{q}}) &=\phantom{+}
           \sum_{n,a}\frac{1}{\sd{J}^{eijk}}\sd{D}^{in} \sd{g}_{a1}^{enjk} \sd{f}_{ab}^{enjk}(\vec{\sd{q}}) \\
  &\quad + \sum_{n,a}\frac{1}{\sd{J}^{eijk}}\sd{D}^{jn} \sd{g}_{a2}^{eink} \sd{f}_{ab}^{eink}(\vec{\sd{q}}) \\
  &\quad + \sum_{n,a}\frac{1}{\sd{J}^{eijk}}\sd{D}^{kn} \sd{g}_{a3}^{eijn} \sd{f}_{ab}^{eijn}(\vec{\sd{q}}),
\end{split}
\end{equation}
where \(D\) is the element differentiation matrix and \(\sd{J}\) and
\(\sd{g}\) are geometric factors related to the Jacobian of the
transformation from the reference to the physical elements. The computation
of the volume term is most expensive aspect of the entire solver. A visual
impression of the per-element ($e$) loop (and data) layout
for this operation is given in Figure~\ref{fig:stencil}.

\section{Transforming the Computation}
\label{sec:transformation}
We begin our consideration with a program that calculates the flux on each
element in the $r$-reference direction. The program enters the transformation
system described as a short subroutine in Fortran~\cite{kloeckner_loopy_2015}:
\begin{lstlisting}[style=customfortran,linerange=START_R_KERNEL-END_R_KERNEL]
! START_R_KERNEL
subroutine refFluxR(Ne, Nq, geo, D, q, rhsq, p_p0, &
    p_Gamma, p_R)
  implicit none
  integer*4 Ne, Nq, e, i, j, k, n
  real geo(Nq,Nq,Nq,11,Ne), D(Nq,Nq)
  real rhsq(Nq,Nq,Nq,8,Ne), q(Nq,Nq,Nq,8,Ne)

  real U1, U2, U3, Rh, Th, P, Q1, Q2, Q3
  real U1flx, U2flx, U3flx, Rhflx, Thflx
  real Q1flx, Q2flx, Q3flx, p_p0, p_Gamma, p_R
  real g11, g21, g31, Jinv, udotGradR, JiD

  do e = 1, Ne
    do k = 1, Nq
      do j = 1, Nq
        do i = 1, Nq
          do n = 1, Nq
!$loopy begin tagged: local_prep
            U1  = q(n,j,k,1,e)
            U2  = q(n,j,k,2,e)
            U3  = q(n,j,k,3,e)
            Rh  = q(n,j,k,4,e)
            Th  = q(n,j,k,5,e)
            Q1  = q(n,j,k,6,e)
            Q2  = q(n,j,k,7,e)
            Q3  = q(n,j,k,8,e)

            g11  = geo(n,j,k,1,e)
            g21  = geo(n,j,k,2,e)
            g31  = geo(n,j,k,3,e)

            Jinv = geo(i,j,k,10,e)

            P = p_p0*(p_R*Th/p_p0) ** p_Gamma
            udotGradR = (g11*U1 + g21*U2 + g31*U3)/Rh
!$loopy end tagged: local_prep

            JiD = Jinv*D(i,n)

!$loopy begin tagged: compute_fluxes
            U1flx = U1*udotGradR + g11*P
            U2flx = U2*udotGradR + g21*P
            U3flx = U3*udotGradR + g31*P
            Rhflx = Rh*udotGradR
            Thflx = Th*udotGradR
            Q1flx = Q1*udotGradR
            Q2flx = Q2*udotGradR
            Q3flx = Q3*udotGradR
!$loopy end tagged: compute_fluxes

            rhsq(i,j,k,1,e) = rhsq(i,j,k,1,e) - JiD*U1flx
            rhsq(i,j,k,2,e) = rhsq(i,j,k,2,e) - JiD*U2flx
            rhsq(i,j,k,3,e) = rhsq(i,j,k,3,e) - JiD*U3flx
            rhsq(i,j,k,4,e) = rhsq(i,j,k,4,e) - JiD*Rhflx
            rhsq(i,j,k,5,e) = rhsq(i,j,k,5,e) - JiD*Thflx
            rhsq(i,j,k,6,e) = rhsq(i,j,k,6,e) - JiD*Q1flx
            rhsq(i,j,k,7,e) = rhsq(i,j,k,7,e) - JiD*Q2flx
            rhsq(i,j,k,8,e) = rhsq(i,j,k,8,e) - JiD*Q3flx
          end do
        end do
      end do
    end do
  end do
end subroutine refFluxR
! END_R_KERNEL

subroutine refFluxS(Ne, Nq, geo, D, q, rhsq, p_p0, &
    p_Gamma, p_R)
  implicit none

  integer*4 Ne, Nq, e, i, j, k, n
  real geo(Nq, Nq, Nq, 11, Ne)
  real p_p0, p_Gamma, p_R
  real D(Nq,Nq)
  real q(Nq, Nq, Nq, 8, Ne)
  real rhsq(Nq, Nq, Nq, 8, Ne)

  real U1, U2, U3, Rh, Th, P, Q1, Q2, Q3
  real U1flx, U2flx, U3flx, Rhflx, Thflx, Q1flx, Q2flx, Q3flx
  real g12, g22, g32, Jinv, udotGradS, JiD

  do e = 1, Ne
    do k = 1, Nq
      do j = 1, Nq
        do i = 1, Nq
          do n = 1, Nq

!$loopy begin tagged: local_prep
! START_S_INPUT
            U1  = q(i,n,k,1,e)
            U2  = q(i,n,k,2,e)
            U3  = q(i,n,k,3,e)
! END_S_INPUT
            Rh  = q(i,n,k,4,e)
            Th  = q(i,n,k,5,e)
            Q1  = q(i,n,k,6,e)
            Q2  = q(i,n,k,7,e)
            Q3  = q(i,n,k,8,e)

            g12  = geo(i,n,k,4,e)
            g22  = geo(i,n,k,5,e)
            g32  = geo(i,n,k,6,e)

            Jinv = geo(i,j,k,10,e)

            P = p_p0*(p_R*Th/p_p0) ** p_Gamma
            udotGradS = (g12*U1 + g22*U2 + g32*U3)/Rh
!$loopy end tagged: local_prep

! START_S_DIFF
            JiD = Jinv*D(j,n)
! END_S_DIFF

!$loopy begin tagged: compute_fluxes
            U1flx = U1*udotGradS + g12*P
            U2flx = U2*udotGradS + g22*P
            U3flx = U3*udotGradS + g32*P
            Rhflx = Rh*udotGradS
            Thflx = Th*udotGradS
            Q1flx = Q1*udotGradS
            Q2flx = Q2*udotGradS
            Q3flx = Q3*udotGradS
!$loopy end tagged: compute_fluxes

            rhsq(i, j, k, 1, e) = rhsq(i, j, k, 1, e) - JiD*U1flx
            rhsq(i, j, k, 2, e) = rhsq(i, j, k, 2, e) - JiD*U2flx
            rhsq(i, j, k, 3, e) = rhsq(i, j, k, 3, e) - JiD*U3flx
            rhsq(i, j, k, 4, e) = rhsq(i, j, k, 4, e) - JiD*Rhflx
            rhsq(i, j, k, 5, e) = rhsq(i, j, k, 5, e) - JiD*Thflx
            rhsq(i, j, k, 6, e) = rhsq(i, j, k, 6, e) - JiD*Q1flx
            rhsq(i, j, k, 7, e) = rhsq(i, j, k, 7, e) - JiD*Q2flx
            rhsq(i, j, k, 8, e) = rhsq(i, j, k, 8, e) - JiD*Q3flx
          end do
        end do
      end do
    end do
  end do
end subroutine refFluxS

!$loopy begin
! try:
!     opt_level
! except NameError:
!     opt_level = 8
!
! try:
!     Nq
! except NameError:
!     Nq = 8
!$loopy end

! #START_XFORM_SETUP
!$loopy begin
! from loopy import *
! import loopy.match as m
!
! f_r, f_s = parse_fortran(SOURCE, FILENAME, auto_dependencies=False)
! #END_XFORM_SETUP
!
! #START_XFORM_FUSE
! f = fuse_kernels([f_r, f_s], suffixes=["_r", "_s"])
! #END_XFORM_FUSE
!
! #START_XFORM_PRELIMINARIES
! f = fix_parameters(f, Nq=Nq)
! f = fix_parameters(f, p_p0=1, p_Gamma=1.4, p_R=1)
! f = assume(f, "Ne >= 1")
! #END_XFORM_PRELIMINARIES
!
! ref_f = f
!
! if opt_level == 0:
!     RESULT = [ref_f, f]
!
! #START_XFORM_LOOPS
! f = set_loop_priority(f, "k")
! f = tag_inames(f, "e:g.0, i:l.0, j:l.1")
! #END_XFORM_LOOPS
!
! if opt_level == 1:
!     RESULT = [ref_f, f]
!
! #START_XFORM_DATA_LAYOUT
! f = tag_data_axes(f, "D", "N0,N1")
!
! for name in ["q", "rhsq"]:
!   f = set_array_dim_names(f, name, "i,j,k,field,e")
!
!   f = split_array_dim(
!       f, (name, 3, "F"), 4, auto_split_inames=False)
!   f = tag_data_axes(f, name, "N0,N1,N2,vec,N4,N3")
! #END_XFORM_DATA_LAYOUT
!
! ref_f = f
! if opt_level == 2:
!     RESULT = [ref_f, f]
!
! #START_XFORM_FETCH_D
! f = add_prefetch(f, "D[:,:]")
! #END_XFORM_FETCH_D
!
! if opt_level == 3:
!     RESULT = [ref_f, f]
!
! #START_XFORM_LOCAL_PREP_SUBST
! local_prep_var_names = set()
! for insn in find_instructions(f, "tag:local_prep"):
!     assignee = insn.assignee_name
!     local_prep_var_names.add(assignee)
!     f = assignment_to_subst(f, assignee)
!
! f = assignment_to_subst(f, "JiD_r")
! f = assignment_to_subst(f, "JiD_s")
! #END_XFORM_LOCAL_PREP_SUBST
!
! if opt_level == 4:
!     RESULT = [ref_f, f]
!
! #START_XFORM_FLUX_TEMP
! comps = ["U1", "U2", "U3", "Rh", "Th", "Q1", "Q2", "Q3"]
!
! for comp in comps:
!   for ref_ax, flux_inames, flux_precomp_inames in [
!         ("r", ("j", "n",), ("jj","ii",)),
!         ("s", ("i", "n",), ("ii","jj",)),
!         ]:
!     flux_var = comp+"flx_"+ref_ax
!     f = assignment_to_subst(f, flux_var)
!
!     flux_store_name = "flux_store_" + comp + ref_ax
!
!     f = precompute(f, flux_var+"_subst", flux_inames,
!       temporary_name=flux_store_name,
!       precompute_inames=flux_precomp_inames,
!       default_tag=None)
!     f = tag_data_axes(
!       f, flux_store_name, {"ii":"N0", "jj":"N1"})
!
! #END_XFORM_FLUX_TEMP
!
! #START_XFORM_FLUX_TMPUSAGE
! f = tag_inames(f, "ii:l.0,jj:l.1")
!
! for ref_ax in ["r", "s"]:
!     f = alias_temporaries(f,
!       ["flux_store_" + comp + ref_ax for comp in comps])
! #END_XFORM_FLUX_TMPUSAGE
!
! #START_XFORM_RENAME_N
! for comp in comps:
!   for ref_ax in ["r", "s"]:
!     flux_store_name = "flux_store_" + comp + ref_ax
!     f = rename_iname(f, "n", "n_"+comp,
!       within=m.Reads(flux_store_name), existing_ok=True)
! #END_XFORM_RENAME_N
!
! if opt_level == 5:
!     RESULT = [ref_f, f]
!
! #START_XFORM_PREP_PRECOMPUTE
! for prep_vname in local_prep_var_names:
!     if prep_vname.startswith("Jinv") or "_s" in prep_vname:
!         continue
!     f = precompute(f,
!       find_one_rule_matching(f, prep_vname+"_*subst*"))
! #END_XFORM_PREP_PRECOMPUTE
!
! if opt_level == 6:
!     RESULT = [ref_f, f]
!
! #START_XFORM_Q_PREFETCH
! f = add_prefetch(f, "q[ii,jj,k,:,:,e]")
! #END_XFORM_Q_PREFETCH
! #START_XFORM_RHSQ_BUFFER
! f = buffer_array(f, "rhsq", (),
!       fetch_bounding_box=True, default_tag="for",
!       init_expression="0",
!       store_expression="base + buffer")
! #END_XFORM_RHSQ_BUFFER
!
! #START_XFORM_VECTOR_ACCESS
! f = tag_data_axes(f, "rhsq_buf", "vec,N0")
! f = tag_data_axes(f, "q_fetch", "vec,N0")
! f = tag_data_axes(f, "D_fetch", "N0,N1")
! f = tag_inames(f,
!  "rhsq_init_field_inner:vec,rhsq_store_field_inner:vec,"
!  "rhsq_init_field_outer:unr,rhsq_store_field_outer:unr,"
!  "q_dim_field_outer:unr,q_dim_field_inner:vec")
! #END_XFORM_VECTOR_ACCESS
!
! if opt_level == 7:
!     RESULT = [ref_f, f]
!
! #START_XFORM_DISTRIBUTIVE
! f = collect_common_factors_on_increment(f, "rhsq_buf")
! #END_XFORM_DISTRIBUTIVE
!
! if opt_level == 8:
!     RESULT = [ref_f, f]
!
!$loopy end
\end{lstlisting}

\noindent
In many ways, the above represents a `typical' Fortran code. It represents a
(more or less) direct translation of the first term on the right-hand
side of the volume discretization term~\eqref{volterm},
faithful to Fortran's original spirit of
`formula translation'. A deep nest of \texttt{do} loops dominates the
structure. Nonetheless, a few peculiar aspects are of note: first, the code is written
entirely without regard to performance. With reasonable clarity, it exhibits
the computational intent. Second, a small amount of annotation is visible. Two
regions, delimited by \texttt{loopy begin/end tagged} markers are shown. These
carry no meaning of their own. They merely serve as location markers for
subsequent transformation operations.

One common performance challenge that stands in direct competition with code
clarity is reuse of data and intermediate results across multiple calculations.
In the context of this example, a salient instance of this issue is the
computation of fluxes in multiple reference directions. All such calculations
refer to the same input data, compute the same (e.g. \verb|P|) or related (e.g. \verb|...flx|) intermediate results.
Yet, a subroutine that merges a number of such calculations invariably
more cluttered and less transparent than the code shown above.

To illustrate, we include in our consideration a further computational
subroutine to compute the fluxes in the $s$-reference direction. This routine
differs from the above merely in the data access pattern. Specifically, the
input of the field variables uses indices as follows:
\begin{lstlisting}[style=customfortran,linerange=START_S_INPUT-END_S_INPUT]
! START_R_KERNEL
subroutine refFluxR(Ne, Nq, geo, D, q, rhsq, p_p0, &
    p_Gamma, p_R)
  implicit none
  integer*4 Ne, Nq, e, i, j, k, n
  real geo(Nq,Nq,Nq,11,Ne), D(Nq,Nq)
  real rhsq(Nq,Nq,Nq,8,Ne), q(Nq,Nq,Nq,8,Ne)

  real U1, U2, U3, Rh, Th, P, Q1, Q2, Q3
  real U1flx, U2flx, U3flx, Rhflx, Thflx
  real Q1flx, Q2flx, Q3flx, p_p0, p_Gamma, p_R
  real g11, g21, g31, Jinv, udotGradR, JiD

  do e = 1, Ne
    do k = 1, Nq
      do j = 1, Nq
        do i = 1, Nq
          do n = 1, Nq
!$loopy begin tagged: local_prep
            U1  = q(n,j,k,1,e)
            U2  = q(n,j,k,2,e)
            U3  = q(n,j,k,3,e)
            Rh  = q(n,j,k,4,e)
            Th  = q(n,j,k,5,e)
            Q1  = q(n,j,k,6,e)
            Q2  = q(n,j,k,7,e)
            Q3  = q(n,j,k,8,e)

            g11  = geo(n,j,k,1,e)
            g21  = geo(n,j,k,2,e)
            g31  = geo(n,j,k,3,e)

            Jinv = geo(i,j,k,10,e)

            P = p_p0*(p_R*Th/p_p0) ** p_Gamma
            udotGradR = (g11*U1 + g21*U2 + g31*U3)/Rh
!$loopy end tagged: local_prep

            JiD = Jinv*D(i,n)

!$loopy begin tagged: compute_fluxes
            U1flx = U1*udotGradR + g11*P
            U2flx = U2*udotGradR + g21*P
            U3flx = U3*udotGradR + g31*P
            Rhflx = Rh*udotGradR
            Thflx = Th*udotGradR
            Q1flx = Q1*udotGradR
            Q2flx = Q2*udotGradR
            Q3flx = Q3*udotGradR
!$loopy end tagged: compute_fluxes

            rhsq(i,j,k,1,e) = rhsq(i,j,k,1,e) - JiD*U1flx
            rhsq(i,j,k,2,e) = rhsq(i,j,k,2,e) - JiD*U2flx
            rhsq(i,j,k,3,e) = rhsq(i,j,k,3,e) - JiD*U3flx
            rhsq(i,j,k,4,e) = rhsq(i,j,k,4,e) - JiD*Rhflx
            rhsq(i,j,k,5,e) = rhsq(i,j,k,5,e) - JiD*Thflx
            rhsq(i,j,k,6,e) = rhsq(i,j,k,6,e) - JiD*Q1flx
            rhsq(i,j,k,7,e) = rhsq(i,j,k,7,e) - JiD*Q2flx
            rhsq(i,j,k,8,e) = rhsq(i,j,k,8,e) - JiD*Q3flx
          end do
        end do
      end do
    end do
  end do
end subroutine refFluxR
! END_R_KERNEL

subroutine refFluxS(Ne, Nq, geo, D, q, rhsq, p_p0, &
    p_Gamma, p_R)
  implicit none

  integer*4 Ne, Nq, e, i, j, k, n
  real geo(Nq, Nq, Nq, 11, Ne)
  real p_p0, p_Gamma, p_R
  real D(Nq,Nq)
  real q(Nq, Nq, Nq, 8, Ne)
  real rhsq(Nq, Nq, Nq, 8, Ne)

  real U1, U2, U3, Rh, Th, P, Q1, Q2, Q3
  real U1flx, U2flx, U3flx, Rhflx, Thflx, Q1flx, Q2flx, Q3flx
  real g12, g22, g32, Jinv, udotGradS, JiD

  do e = 1, Ne
    do k = 1, Nq
      do j = 1, Nq
        do i = 1, Nq
          do n = 1, Nq

!$loopy begin tagged: local_prep
! START_S_INPUT
            U1  = q(i,n,k,1,e)
            U2  = q(i,n,k,2,e)
            U3  = q(i,n,k,3,e)
! END_S_INPUT
            Rh  = q(i,n,k,4,e)
            Th  = q(i,n,k,5,e)
            Q1  = q(i,n,k,6,e)
            Q2  = q(i,n,k,7,e)
            Q3  = q(i,n,k,8,e)

            g12  = geo(i,n,k,4,e)
            g22  = geo(i,n,k,5,e)
            g32  = geo(i,n,k,6,e)

            Jinv = geo(i,j,k,10,e)

            P = p_p0*(p_R*Th/p_p0) ** p_Gamma
            udotGradS = (g12*U1 + g22*U2 + g32*U3)/Rh
!$loopy end tagged: local_prep

! START_S_DIFF
            JiD = Jinv*D(j,n)
! END_S_DIFF

!$loopy begin tagged: compute_fluxes
            U1flx = U1*udotGradS + g12*P
            U2flx = U2*udotGradS + g22*P
            U3flx = U3*udotGradS + g32*P
            Rhflx = Rh*udotGradS
            Thflx = Th*udotGradS
            Q1flx = Q1*udotGradS
            Q2flx = Q2*udotGradS
            Q3flx = Q3*udotGradS
!$loopy end tagged: compute_fluxes

            rhsq(i, j, k, 1, e) = rhsq(i, j, k, 1, e) - JiD*U1flx
            rhsq(i, j, k, 2, e) = rhsq(i, j, k, 2, e) - JiD*U2flx
            rhsq(i, j, k, 3, e) = rhsq(i, j, k, 3, e) - JiD*U3flx
            rhsq(i, j, k, 4, e) = rhsq(i, j, k, 4, e) - JiD*Rhflx
            rhsq(i, j, k, 5, e) = rhsq(i, j, k, 5, e) - JiD*Thflx
            rhsq(i, j, k, 6, e) = rhsq(i, j, k, 6, e) - JiD*Q1flx
            rhsq(i, j, k, 7, e) = rhsq(i, j, k, 7, e) - JiD*Q2flx
            rhsq(i, j, k, 8, e) = rhsq(i, j, k, 8, e) - JiD*Q3flx
          end do
        end do
      end do
    end do
  end do
end subroutine refFluxS

!$loopy begin
! try:
!     opt_level
! except NameError:
!     opt_level = 8
!
! try:
!     Nq
! except NameError:
!     Nq = 8
!$loopy end

! #START_XFORM_SETUP
!$loopy begin
! from loopy import *
! import loopy.match as m
!
! f_r, f_s = parse_fortran(SOURCE, FILENAME, auto_dependencies=False)
! #END_XFORM_SETUP
!
! #START_XFORM_FUSE
! f = fuse_kernels([f_r, f_s], suffixes=["_r", "_s"])
! #END_XFORM_FUSE
!
! #START_XFORM_PRELIMINARIES
! f = fix_parameters(f, Nq=Nq)
! f = fix_parameters(f, p_p0=1, p_Gamma=1.4, p_R=1)
! f = assume(f, "Ne >= 1")
! #END_XFORM_PRELIMINARIES
!
! ref_f = f
!
! if opt_level == 0:
!     RESULT = [ref_f, f]
!
! #START_XFORM_LOOPS
! f = set_loop_priority(f, "k")
! f = tag_inames(f, "e:g.0, i:l.0, j:l.1")
! #END_XFORM_LOOPS
!
! if opt_level == 1:
!     RESULT = [ref_f, f]
!
! #START_XFORM_DATA_LAYOUT
! f = tag_data_axes(f, "D", "N0,N1")
!
! for name in ["q", "rhsq"]:
!   f = set_array_dim_names(f, name, "i,j,k,field,e")
!
!   f = split_array_dim(
!       f, (name, 3, "F"), 4, auto_split_inames=False)
!   f = tag_data_axes(f, name, "N0,N1,N2,vec,N4,N3")
! #END_XFORM_DATA_LAYOUT
!
! ref_f = f
! if opt_level == 2:
!     RESULT = [ref_f, f]
!
! #START_XFORM_FETCH_D
! f = add_prefetch(f, "D[:,:]")
! #END_XFORM_FETCH_D
!
! if opt_level == 3:
!     RESULT = [ref_f, f]
!
! #START_XFORM_LOCAL_PREP_SUBST
! local_prep_var_names = set()
! for insn in find_instructions(f, "tag:local_prep"):
!     assignee = insn.assignee_name
!     local_prep_var_names.add(assignee)
!     f = assignment_to_subst(f, assignee)
!
! f = assignment_to_subst(f, "JiD_r")
! f = assignment_to_subst(f, "JiD_s")
! #END_XFORM_LOCAL_PREP_SUBST
!
! if opt_level == 4:
!     RESULT = [ref_f, f]
!
! #START_XFORM_FLUX_TEMP
! comps = ["U1", "U2", "U3", "Rh", "Th", "Q1", "Q2", "Q3"]
!
! for comp in comps:
!   for ref_ax, flux_inames, flux_precomp_inames in [
!         ("r", ("j", "n",), ("jj","ii",)),
!         ("s", ("i", "n",), ("ii","jj",)),
!         ]:
!     flux_var = comp+"flx_"+ref_ax
!     f = assignment_to_subst(f, flux_var)
!
!     flux_store_name = "flux_store_" + comp + ref_ax
!
!     f = precompute(f, flux_var+"_subst", flux_inames,
!       temporary_name=flux_store_name,
!       precompute_inames=flux_precomp_inames,
!       default_tag=None)
!     f = tag_data_axes(
!       f, flux_store_name, {"ii":"N0", "jj":"N1"})
!
! #END_XFORM_FLUX_TEMP
!
! #START_XFORM_FLUX_TMPUSAGE
! f = tag_inames(f, "ii:l.0,jj:l.1")
!
! for ref_ax in ["r", "s"]:
!     f = alias_temporaries(f,
!       ["flux_store_" + comp + ref_ax for comp in comps])
! #END_XFORM_FLUX_TMPUSAGE
!
! #START_XFORM_RENAME_N
! for comp in comps:
!   for ref_ax in ["r", "s"]:
!     flux_store_name = "flux_store_" + comp + ref_ax
!     f = rename_iname(f, "n", "n_"+comp,
!       within=m.Reads(flux_store_name), existing_ok=True)
! #END_XFORM_RENAME_N
!
! if opt_level == 5:
!     RESULT = [ref_f, f]
!
! #START_XFORM_PREP_PRECOMPUTE
! for prep_vname in local_prep_var_names:
!     if prep_vname.startswith("Jinv") or "_s" in prep_vname:
!         continue
!     f = precompute(f,
!       find_one_rule_matching(f, prep_vname+"_*subst*"))
! #END_XFORM_PREP_PRECOMPUTE
!
! if opt_level == 6:
!     RESULT = [ref_f, f]
!
! #START_XFORM_Q_PREFETCH
! f = add_prefetch(f, "q[ii,jj,k,:,:,e]")
! #END_XFORM_Q_PREFETCH
! #START_XFORM_RHSQ_BUFFER
! f = buffer_array(f, "rhsq", (),
!       fetch_bounding_box=True, default_tag="for",
!       init_expression="0",
!       store_expression="base + buffer")
! #END_XFORM_RHSQ_BUFFER
!
! #START_XFORM_VECTOR_ACCESS
! f = tag_data_axes(f, "rhsq_buf", "vec,N0")
! f = tag_data_axes(f, "q_fetch", "vec,N0")
! f = tag_data_axes(f, "D_fetch", "N0,N1")
! f = tag_inames(f,
!  "rhsq_init_field_inner:vec,rhsq_store_field_inner:vec,"
!  "rhsq_init_field_outer:unr,rhsq_store_field_outer:unr,"
!  "q_dim_field_outer:unr,q_dim_field_inner:vec")
! #END_XFORM_VECTOR_ACCESS
!
! if opt_level == 7:
!     RESULT = [ref_f, f]
!
! #START_XFORM_DISTRIBUTIVE
! f = collect_common_factors_on_increment(f, "rhsq_buf")
! #END_XFORM_DISTRIBUTIVE
!
! if opt_level == 8:
!     RESULT = [ref_f, f]
!
!$loopy end
\end{lstlisting}
(further accesses proceed analogously)
and the differentiation matrix is accessed as:
\begin{lstlisting}[style=customfortran,linerange=START_S_DIFF-END_S_DIFF]
! START_R_KERNEL
subroutine refFluxR(Ne, Nq, geo, D, q, rhsq, p_p0, &
    p_Gamma, p_R)
  implicit none
  integer*4 Ne, Nq, e, i, j, k, n
  real geo(Nq,Nq,Nq,11,Ne), D(Nq,Nq)
  real rhsq(Nq,Nq,Nq,8,Ne), q(Nq,Nq,Nq,8,Ne)

  real U1, U2, U3, Rh, Th, P, Q1, Q2, Q3
  real U1flx, U2flx, U3flx, Rhflx, Thflx
  real Q1flx, Q2flx, Q3flx, p_p0, p_Gamma, p_R
  real g11, g21, g31, Jinv, udotGradR, JiD

  do e = 1, Ne
    do k = 1, Nq
      do j = 1, Nq
        do i = 1, Nq
          do n = 1, Nq
!$loopy begin tagged: local_prep
            U1  = q(n,j,k,1,e)
            U2  = q(n,j,k,2,e)
            U3  = q(n,j,k,3,e)
            Rh  = q(n,j,k,4,e)
            Th  = q(n,j,k,5,e)
            Q1  = q(n,j,k,6,e)
            Q2  = q(n,j,k,7,e)
            Q3  = q(n,j,k,8,e)

            g11  = geo(n,j,k,1,e)
            g21  = geo(n,j,k,2,e)
            g31  = geo(n,j,k,3,e)

            Jinv = geo(i,j,k,10,e)

            P = p_p0*(p_R*Th/p_p0) ** p_Gamma
            udotGradR = (g11*U1 + g21*U2 + g31*U3)/Rh
!$loopy end tagged: local_prep

            JiD = Jinv*D(i,n)

!$loopy begin tagged: compute_fluxes
            U1flx = U1*udotGradR + g11*P
            U2flx = U2*udotGradR + g21*P
            U3flx = U3*udotGradR + g31*P
            Rhflx = Rh*udotGradR
            Thflx = Th*udotGradR
            Q1flx = Q1*udotGradR
            Q2flx = Q2*udotGradR
            Q3flx = Q3*udotGradR
!$loopy end tagged: compute_fluxes

            rhsq(i,j,k,1,e) = rhsq(i,j,k,1,e) - JiD*U1flx
            rhsq(i,j,k,2,e) = rhsq(i,j,k,2,e) - JiD*U2flx
            rhsq(i,j,k,3,e) = rhsq(i,j,k,3,e) - JiD*U3flx
            rhsq(i,j,k,4,e) = rhsq(i,j,k,4,e) - JiD*Rhflx
            rhsq(i,j,k,5,e) = rhsq(i,j,k,5,e) - JiD*Thflx
            rhsq(i,j,k,6,e) = rhsq(i,j,k,6,e) - JiD*Q1flx
            rhsq(i,j,k,7,e) = rhsq(i,j,k,7,e) - JiD*Q2flx
            rhsq(i,j,k,8,e) = rhsq(i,j,k,8,e) - JiD*Q3flx
          end do
        end do
      end do
    end do
  end do
end subroutine refFluxR
! END_R_KERNEL

subroutine refFluxS(Ne, Nq, geo, D, q, rhsq, p_p0, &
    p_Gamma, p_R)
  implicit none

  integer*4 Ne, Nq, e, i, j, k, n
  real geo(Nq, Nq, Nq, 11, Ne)
  real p_p0, p_Gamma, p_R
  real D(Nq,Nq)
  real q(Nq, Nq, Nq, 8, Ne)
  real rhsq(Nq, Nq, Nq, 8, Ne)

  real U1, U2, U3, Rh, Th, P, Q1, Q2, Q3
  real U1flx, U2flx, U3flx, Rhflx, Thflx, Q1flx, Q2flx, Q3flx
  real g12, g22, g32, Jinv, udotGradS, JiD

  do e = 1, Ne
    do k = 1, Nq
      do j = 1, Nq
        do i = 1, Nq
          do n = 1, Nq

!$loopy begin tagged: local_prep
! START_S_INPUT
            U1  = q(i,n,k,1,e)
            U2  = q(i,n,k,2,e)
            U3  = q(i,n,k,3,e)
! END_S_INPUT
            Rh  = q(i,n,k,4,e)
            Th  = q(i,n,k,5,e)
            Q1  = q(i,n,k,6,e)
            Q2  = q(i,n,k,7,e)
            Q3  = q(i,n,k,8,e)

            g12  = geo(i,n,k,4,e)
            g22  = geo(i,n,k,5,e)
            g32  = geo(i,n,k,6,e)

            Jinv = geo(i,j,k,10,e)

            P = p_p0*(p_R*Th/p_p0) ** p_Gamma
            udotGradS = (g12*U1 + g22*U2 + g32*U3)/Rh
!$loopy end tagged: local_prep

! START_S_DIFF
            JiD = Jinv*D(j,n)
! END_S_DIFF

!$loopy begin tagged: compute_fluxes
            U1flx = U1*udotGradS + g12*P
            U2flx = U2*udotGradS + g22*P
            U3flx = U3*udotGradS + g32*P
            Rhflx = Rh*udotGradS
            Thflx = Th*udotGradS
            Q1flx = Q1*udotGradS
            Q2flx = Q2*udotGradS
            Q3flx = Q3*udotGradS
!$loopy end tagged: compute_fluxes

            rhsq(i, j, k, 1, e) = rhsq(i, j, k, 1, e) - JiD*U1flx
            rhsq(i, j, k, 2, e) = rhsq(i, j, k, 2, e) - JiD*U2flx
            rhsq(i, j, k, 3, e) = rhsq(i, j, k, 3, e) - JiD*U3flx
            rhsq(i, j, k, 4, e) = rhsq(i, j, k, 4, e) - JiD*Rhflx
            rhsq(i, j, k, 5, e) = rhsq(i, j, k, 5, e) - JiD*Thflx
            rhsq(i, j, k, 6, e) = rhsq(i, j, k, 6, e) - JiD*Q1flx
            rhsq(i, j, k, 7, e) = rhsq(i, j, k, 7, e) - JiD*Q2flx
            rhsq(i, j, k, 8, e) = rhsq(i, j, k, 8, e) - JiD*Q3flx
          end do
        end do
      end do
    end do
  end do
end subroutine refFluxS

!$loopy begin
! try:
!     opt_level
! except NameError:
!     opt_level = 8
!
! try:
!     Nq
! except NameError:
!     Nq = 8
!$loopy end

! #START_XFORM_SETUP
!$loopy begin
! from loopy import *
! import loopy.match as m
!
! f_r, f_s = parse_fortran(SOURCE, FILENAME, auto_dependencies=False)
! #END_XFORM_SETUP
!
! #START_XFORM_FUSE
! f = fuse_kernels([f_r, f_s], suffixes=["_r", "_s"])
! #END_XFORM_FUSE
!
! #START_XFORM_PRELIMINARIES
! f = fix_parameters(f, Nq=Nq)
! f = fix_parameters(f, p_p0=1, p_Gamma=1.4, p_R=1)
! f = assume(f, "Ne >= 1")
! #END_XFORM_PRELIMINARIES
!
! ref_f = f
!
! if opt_level == 0:
!     RESULT = [ref_f, f]
!
! #START_XFORM_LOOPS
! f = set_loop_priority(f, "k")
! f = tag_inames(f, "e:g.0, i:l.0, j:l.1")
! #END_XFORM_LOOPS
!
! if opt_level == 1:
!     RESULT = [ref_f, f]
!
! #START_XFORM_DATA_LAYOUT
! f = tag_data_axes(f, "D", "N0,N1")
!
! for name in ["q", "rhsq"]:
!   f = set_array_dim_names(f, name, "i,j,k,field,e")
!
!   f = split_array_dim(
!       f, (name, 3, "F"), 4, auto_split_inames=False)
!   f = tag_data_axes(f, name, "N0,N1,N2,vec,N4,N3")
! #END_XFORM_DATA_LAYOUT
!
! ref_f = f
! if opt_level == 2:
!     RESULT = [ref_f, f]
!
! #START_XFORM_FETCH_D
! f = add_prefetch(f, "D[:,:]")
! #END_XFORM_FETCH_D
!
! if opt_level == 3:
!     RESULT = [ref_f, f]
!
! #START_XFORM_LOCAL_PREP_SUBST
! local_prep_var_names = set()
! for insn in find_instructions(f, "tag:local_prep"):
!     assignee = insn.assignee_name
!     local_prep_var_names.add(assignee)
!     f = assignment_to_subst(f, assignee)
!
! f = assignment_to_subst(f, "JiD_r")
! f = assignment_to_subst(f, "JiD_s")
! #END_XFORM_LOCAL_PREP_SUBST
!
! if opt_level == 4:
!     RESULT = [ref_f, f]
!
! #START_XFORM_FLUX_TEMP
! comps = ["U1", "U2", "U3", "Rh", "Th", "Q1", "Q2", "Q3"]
!
! for comp in comps:
!   for ref_ax, flux_inames, flux_precomp_inames in [
!         ("r", ("j", "n",), ("jj","ii",)),
!         ("s", ("i", "n",), ("ii","jj",)),
!         ]:
!     flux_var = comp+"flx_"+ref_ax
!     f = assignment_to_subst(f, flux_var)
!
!     flux_store_name = "flux_store_" + comp + ref_ax
!
!     f = precompute(f, flux_var+"_subst", flux_inames,
!       temporary_name=flux_store_name,
!       precompute_inames=flux_precomp_inames,
!       default_tag=None)
!     f = tag_data_axes(
!       f, flux_store_name, {"ii":"N0", "jj":"N1"})
!
! #END_XFORM_FLUX_TEMP
!
! #START_XFORM_FLUX_TMPUSAGE
! f = tag_inames(f, "ii:l.0,jj:l.1")
!
! for ref_ax in ["r", "s"]:
!     f = alias_temporaries(f,
!       ["flux_store_" + comp + ref_ax for comp in comps])
! #END_XFORM_FLUX_TMPUSAGE
!
! #START_XFORM_RENAME_N
! for comp in comps:
!   for ref_ax in ["r", "s"]:
!     flux_store_name = "flux_store_" + comp + ref_ax
!     f = rename_iname(f, "n", "n_"+comp,
!       within=m.Reads(flux_store_name), existing_ok=True)
! #END_XFORM_RENAME_N
!
! if opt_level == 5:
!     RESULT = [ref_f, f]
!
! #START_XFORM_PREP_PRECOMPUTE
! for prep_vname in local_prep_var_names:
!     if prep_vname.startswith("Jinv") or "_s" in prep_vname:
!         continue
!     f = precompute(f,
!       find_one_rule_matching(f, prep_vname+"_*subst*"))
! #END_XFORM_PREP_PRECOMPUTE
!
! if opt_level == 6:
!     RESULT = [ref_f, f]
!
! #START_XFORM_Q_PREFETCH
! f = add_prefetch(f, "q[ii,jj,k,:,:,e]")
! #END_XFORM_Q_PREFETCH
! #START_XFORM_RHSQ_BUFFER
! f = buffer_array(f, "rhsq", (),
!       fetch_bounding_box=True, default_tag="for",
!       init_expression="0",
!       store_expression="base + buffer")
! #END_XFORM_RHSQ_BUFFER
!
! #START_XFORM_VECTOR_ACCESS
! f = tag_data_axes(f, "rhsq_buf", "vec,N0")
! f = tag_data_axes(f, "q_fetch", "vec,N0")
! f = tag_data_axes(f, "D_fetch", "N0,N1")
! f = tag_inames(f,
!  "rhsq_init_field_inner:vec,rhsq_store_field_inner:vec,"
!  "rhsq_init_field_outer:unr,rhsq_store_field_outer:unr,"
!  "q_dim_field_outer:unr,q_dim_field_inner:vec")
! #END_XFORM_VECTOR_ACCESS
!
! if opt_level == 7:
!     RESULT = [ref_f, f]
!
! #START_XFORM_DISTRIBUTIVE
! f = collect_common_factors_on_increment(f, "rhsq_buf")
! #END_XFORM_DISTRIBUTIVE
!
! if opt_level == 8:
!     RESULT = [ref_f, f]
!
!$loopy end
\end{lstlisting}
Another minor difference is that the $s$-kernel uses a different set of geometric factors.
All other aspects, including the final output increments, are exactly the same.

Transformations in \loopy\ are expressed using the Python programming
language.  This code can be given either in a separate location or, as
shown here, as part of the same Fortran file in a directive-like comment
block.

As we proceed through the transformations, we will occasionally denote
intermediate versions of the kernel as being of a certain optimization level.
Where relevant, this is indicated in parentheses as `[Opt.\ level $N$]'.
These optimization levels are then later referenced in the results section to
clarify the individual performance impact of each of the transformation steps.

The transformation begins by receiving the translation unit
(including the comment block) in an implicitly defined variable
\texttt{SOURCE} which it then translates from Fortran into Loopy's
intermediate representation, resulting in two kernels which are returned:

\prexformincl{SETUP}

\noindent
We omit the Fortran comment markers (`\verb|!|') in the following. Next up,
the $r$- and $s$-flux computation kernels are fused together.
This operation joins loop axes (`\emph{inames}') with matching names. The projection of
the domains of each pair of kernels onto their common subset of inames
must match in order for fusion to succeed. Kernel arguments with
identical names are likewise merged, assuming non-conflicting declarations.
Local variables are kept private to each of the fused kernels by applying
a user-given suffix, here `\verb|_r|' and `\verb|_s|'.

\xformincl{FUSE}

\noindent
In situations more complicated than this one, data flow between kernels
may be specified as part of fusion to determine dependencies.

Next, fixing values of parameters helps determine control flow by fixing loop
bounds and removing run-time data references. A non-emptiness assumption
further helps reduce unnecessary conditionals.

\xformincl{PRELIMINARIES}

\noindent
To further determine control flow, we set the ordering of the sequential
loops making the intra-element axis \verb|k| the outer loop.
Furthermore, we assign the element number iname \verb|e| to the
abstract `core' axis and the intra-element axes \verb|i| and \verb|j|
to two of the `SIMD lane' axes exposed in the OpenCL model.

\xformincl{LOOPS}

\optlevel{1}
The following section determines the data layouts of the bulk variables
\verb|q| and \verb|rhsq|.
At the coarsest level, this is done by specifying a ``nesting
order'' of axes, sorting them from fastest- to slowest-varying. This is
indicated by a capital \verb|N| followed by an integer. The computation of the surface
term (mentioned in Section~\ref{sec:comp-overview}) amounts to scattered,
indirect memory access. To better match this access to the wide bus width and
relative dearth of caches on the target hardware, we use short vector types (of
length four) as the basic granularity of our access. We map the `\verb|field|' axis
(of length eight) of the arrays containing our degrees of freedom across the
vector entries. To facilitate this, we split the axis in two as $\mathtt{field}=4\mathtt{field\_outer} + \mathtt{field\_inner}$,
where \verb|field_inner| and \verb|field_outer| are now exposed as separate array
axes arranged in ``column-major'' (i.e., \verb|"F"|ortran) order, i.e.,
\verb|field_inner| comes first. \verb|field_inner|, being of fixed length 4,
is then ready to be tagged for implementation as a short vector (`\verb|"vec"|').

In addition, this section stipulates some identifiers for array axes which are
subsequently used in automatic name generation for, e.g., inames
in precomputation dealing with this array axis.

\xformincl{DATA_LAYOUT}

\optlevel{2}
Since the differentiation matrix \verb|D| is referenced frequently, it makes
sense to bring it into local scratchpad memory, which is accomplished as follows:

\xformincl{FETCH_D}

\optlevel{3}
The main computation in this kernel consists of two layers: first,
the purely local evaluation of the flux function $f_a(\mathbf q)$ followed by
the non-local derivative operation. To allow this multi-layered computation to
be transformed, we employ substitution rules as an intermediate form. These can
be imagined as parametric, expression-level macros that may always be expanded.
One of their main purposes is to assign a name to a certain intermediate result.
The Fortran form of our kernel begins by storing field values into local variables.
This can be seen in an area of the code carrying a `\texttt{loopy begin tagged:
local\_prep}' annotation. This occurs in preparation for the pre-computation
and storing of the flux values, which refer to this data. To retain the
semantic content of these variable names while removing the computational
inconvenience, we use an \verb|assignment_to_subst| transformation to obtain an
appropriate substitution rule. In code, the effect of this is the elimination
of the temporary variable, all references to which are replaced by references
to the rule, which in turn expands to the expression originally assigned to the temporary.
The annotation above is used to create a list of
all the variable names for which this is needed. Going forward, we will refer
to this data as `degree of freedom' data.

\xformincl{LOCAL_PREP_SUBST}

\optlevel{4}
In a similar fashion, we make substitution rules of the flux
computations, for each component of the solution vector. Given these rules, we
use the \verb|precompute| transformation to create a loop nest that computes
and stores the values of the fluxes in a temporary variable. We perform this
precomputation for one $k$-slice at a time (cf.\ Figure~\ref{fig:stencil}),
yielding a family of two-dimensional calculations each determined by two pairs
of two inames. The first such pair determines the access footprint, i.e., the
set of inames that span the index space for which the pre-computation will
replace the actual memory access and subsequent expression evaluation. The
second such pair determines the inames over which the pre-computation is
actually carried out. Since this transformation needs to be carried out once
for the $r$- and once for the $s$-kernel, we encode a (Python) \verb|for| loop. For the
$r$-kernel, the first pair is \verb|(j,n)| (cf.\ the \verb|local_prep| section
of the code). In each case, the pair of inames carrying out the pre-computation
is a permutation of the new inames \verb|ii| and \verb|jj|. We also fix the
data layout for the newly created temporary.

The two main operational aspects of the \verb|precompute| transformation are
the creation of the temporary variable and the rewriting of the input
substitution rule into an assignment operation. The latter mainly involves
changing subscripts to use the inames that are being used to carry out the
pre-computation.

It is worth noting that, since the read operations on our degree of freedom
data exist in a (sub-)substitution rule as created above, the subscript
information therein tracks the changes being made. This is the case because,
for \loopy's purposes, term rewriting is applied as if the substitution rules
did not exist, i.e., term rewriting, at least notionally, always operates on a
`fully expanded' version of the expression tree. Substitution rules are
nonetheless preserved, however it may be the case that multiple versions of a
rule are created depending on different rewritings occurring on a
per-usage-site basis. Similarly, substitution rules with identical content are
merged. Through this merging mechanism as well as an appropriate choice of the
pre-computation inames, we are able to merge the degree-of-freedom reads from
the $r$- and the $s$-kernel, ensuring that the resulting fused kernel only
performs a single read access to that data.

\xformincl{FLUX_TEMP}

\noindent Next, we determine that the new inames \verb|ii| and \verb|jj| should
be mapped to the abstract SIMD lane indices along axis 0 and 1. To conserve
temporary, on-chip storage, we next request that, within each of the $r$- and
the $s$-parts of the fused kernel, the temporary flux storage for each solution
component be aliased, resulting in two storage areas being used, one for the
$r$-kernel and one for the $s$-kernel. This aliasing automatically creates
scheduling constraints that prohibit the range from the first right to the last
read for each temporary to overlap with the same range for another temporary
that is aliased to the same actual storage.

\xformincl{FLUX_TMPUSAGE}

\noindent These scheduling constraints necessarily prevent that a single \verb|n|
(i.e., summation) loop can be used, since at any one time only two sets of
precomputed flux data are available. Further, since \loopy\ uses inames as unique
identifiers of loops, a (kernel-level) \verb|for| loop that has been opened and closed once
cannot be opened again. Therefore, we use the \verb|rename_iname|
transformation to create copies of the iname \verb|n|. A unique copy of the
\verb|n| iname is created for each pair of flux temporaries, to be used by the
summation instructions that refer to them. That set of instructions is
identified by the \verb|Reads(flux_store_name)| match query.

\xformincl{RENAME_N}

\optlevel{5}
To appropriately limit the scope of transformations, such as in the previous scenario,
\loopy\ provides a small, but fairly comprehensive query language to match
program context, based on features of instructions and substitution rules.

Moving along, we once again use the \verb|precompute| transformation, this time
to fetch the degree of freedom data into temporary variables. Note that
\verb|Jinv| is excluded from this pre-computation because it follows a
different access pattern.

\xformincl{PREP_PRECOMPUTE}

\optlevel{6}
Next, by using the \verb|add_prefetch| transformation, we create a batch fetch
process (which will subsequently be vectorized) to load the degree-of-freedom
data across the field components on to the chip. The transformation
\verb|add_prefetch| combines
(1) substitution rule creation based on an existing global (i.e., off-chip)
variable, (2) replacement of appropriate references to that variable with
references to that substitution rule, and (3) the application of the
\verb|precompute| transformation to that same substitution rule. The net effect
of this is for off-chip data to be stored in an on-chip temporary variable and
then referenced from there instead of from global memory.

\xformincl{Q_PREFETCH}

\noindent
The Fortran code, as its last step of processing, performs in-place updates
on the computed time derivative \verb|rhsq| of the degree-of-freedom data. This
particular access, while easy to code, is not efficient, because it
touches each of these off-chip quantities $2N_{q}$ times. The performance issue of
this access is further exacerbated by the chosen (vectorized) memory layout,
leading to non-contiguous access. Buffering this data on the chip for the
duration of the kernel addresses all these performance concerns:

\xformincl{RHSQ_BUFFER}

\optlevel{7}
We fix the data layouts of the on-chip versions of \verb|D|, \verb|q| and
\verb|rhsq|. For the latter two, we ensure that their data format matches the
vectorization of their off-chip versions. To ensure that the access carrying
this data onto and off of the chip proceeds using full width vectors, we also
need to tag the relevant inames with the \verb|vec| iname implementation tag.
Note that the execution recipe as chosen by the iname tag is independent of the
data layout, as data stored as vectors can still be accessed in a scalar
fashion, hence the necessity to specify both.

\xformincl{VECTOR_ACCESS}

\noindent Our last transformation step takes advantage of the distributive law
in the setting where a variable being additively updated is also buffered on
the chip. In this setting, common factors can be `pulled out' and applied only
once at the time when the buffered variable is being written back to its
off-chip location. This is realized by the following transformation:

\xformincl{DISTRIBUTIVE}

\optlevel{8}
This completes our walkthrough of the transformation and yields the final
version of the kernel.

\section{Results}
\label{sec:results}
\begin{table}
  \centering
  \begin{tabular}{@{} S[
                        table-number-alignment = center,
                      ]
                      S[
                        table-number-alignment = center,
                      ]
                      S[
                        table-number-alignment = center,
                      ]
                      S[
                        table-number-alignment = center,
                      ]
                      S[
                        table-number-alignment = center,
                      ]@{}} \toprule
  {Opt.} &  {Wall Time}    & {Speedup}       & {\si{\giga\flops}} & {Bandwidth} \\
  {Level}               &       {(\si{\milli\second})} & & & {(\si{\giga\byte\per\second})} \\ \midrule
    \multicolumn{5}{r}{Radeon R9 FURY X} \\
      1 & 25.01630 &  1.0     &   88.27 &  262.56 \\
      2 & 13.49917 &  1.85317 &  163.59 &  486.57 \\
      3 & 14.15880 &  1.76684 &  155.97 &  447.90 \\
      4 & 13.96988 &  1.79073 &  372.90 & 1524.01 \\
      5 &  2.42899 & 10.2991  &  757.62 & 3053.79 \\
      6 &  2.13116 & 11.7383  &  707.40 & 2643.63 \\
      7 &  1.37435 & 18.2023  & 1117.55 & 1627.40 \\
      8 &  1.36121 & 18.378   &  816.36 &  395.18 \\
    \multicolumn{5}{r}{GeForce GTX TITAN X} \\
      1 &  7.17732 & 1.0      &  307.68 &  915.14 \\
      2 &  5.14222 & 1.39576  &  429.45 & 1277.32 \\
      3 &  5.05879 & 1.41878  &  436.53 & 1253.62 \\
      4 &  9.16962 & 0.782728 &  568.11 & 2321.83 \\
      5 &  1.83405 & 3.91337  & 1003.38 & 4044.40 \\
      6 &  2.08800 & 3.43741  &  722.02 & 2698.27 \\
      7 &  2.03898 & 3.52005  &  753.27 & 1096.93 \\
      8 &  2.06197 & 3.48081  &  538.92 &  260.88 \\
    \multicolumn{5}{r}{Tesla K40c} \\
      1 & 14.29650 & 1.0      & 154.46 &  459.43 \\
      2 & 13.35464 & 1.07053  & 165.36 &  491.84 \\
      3 & 13.28151 & 1.07642  & 166.27 &  477.49 \\
      4 & 20.96802 & 0.681824 & 248.44 & 1015.37 \\
      5 &  4.44362 & 3.21731  & 414.13 & 1669.27 \\
      6 &  3.60599 & 3.96465  & 418.08 & 1562.40 \\
      7 &  3.43723 & 4.15931  & 446.84 &  650.70 \\
      8 &  3.28481 & 4.35231  & 338.29 &  163.76 \\
    \bottomrule
  \end{tabular}
  \caption{Performance of the kernel for the various optimization
    levels described in
    Section~\ref{sec:transformation}.  Note, the \si{\flops} and bandwidth
    are estimated, not measured, as described in
    Section~\ref{sec:results}.}\label{tab:step-perf-table}
\end{table}

\emph{Performance Throughout the Transformation.}
An interesting data set whose collection is enabled by the transformation-based
nature of our work arises from the performance characteristics across all versions
of the kernel.
To this end, we have indicated various points within the preceding section as
`optimization levels.' Table~\ref{tab:step-perf-table} shows this data for three recent GPU
architectures. These results were obtained using \verb|Nq=8| and \verb|Ne=6910|.
We note that even the baseline (level 1) version of the kernel makes full use of the
available parallelism and concurrency on the GPUs. The speedup number
indicated thus summarizes what gains in wall time can be achieved by performing
additional tuning beyond such parallelization.

Two features of this data are immediately striking: first, performance is not
necessarily monotonic. Some transformations may individually lose performance,
but, in combination with later transformations enabled by them, make an even
larger gain possible. Second, some GPU architectures are more sensitive to
specific tuning efforts than others. Specifically, in our example, the AMD
GPU appears to benefit to a much greater extent than the others.

For a complementary point of view, Table~\ref{tab:step-perf-table} also shows
absolute performance numbers in terms of floating-point operation
rate and memory bandwidth. While each computation for which timing data is
displayed computes the same numerical output, it is worth while to observe that
some intermediate optimization levels achieve far higher FLOP rates or memory
bandwidths than the final versions. These intermediate versions are not superior
despite the higher performance because they ultimately take longer to complete
in terms of wall time. This does highlight that some of the
transformations realized by \loopy\ amount to algorithmic changes affecting the
overall complexity of the computation.

%
%
%
%

\begin{table}
  \centering
  \begin{tabular}{@{} S[
                        table-number-alignment = center,
                      ]
                      S[
                        table-number-alignment = center,
                      ]
                      S[
                        table-number-alignment = center,
                      ]
                      S[
                        table-number-alignment = center,
                      ]
                      S[
                        table-number-alignment = center,
                      ]@{}} \toprule
  {\texttt{Nq}} & {\texttt{Ne}} &  {Wall Time}            & {\si{\giga\flops}} & {Bandwidth} \\
                &               &  {(\si{\milli\second})} &                    & {(\si{\giga\byte\per\second})} \\ \midrule
    \multicolumn{5}{r}{Radeon R9 FURY X} \\
      4 &  55296 & 3.19944 &  205.74 & 168.13 \\
      8 &   6912 & 1.36146 &  816.21 & 395.11 \\
     12 &   2048 & 4.26229 &  366.99 & 126.20 \\
     16 &    864 & 6.45935 &  312.29 &  83.28 \\
    \multicolumn{5}{r}{GeForce GTX TITAN X} \\
      4 &  55296 & 1.78229 & 369.33 & 301.81 \\
      8 &   6912 & 2.06701 & 537.60 & 260.24 \\
     12 &   2048 & 2.20707 & 708.73 & 243.73 \\
     16 &    864 & 2.11778 & 952.50 & 254.00 \\
    \multicolumn{5}{r}{Tesla K40c} \\
      4 &  55296 & 5.54140 & 118.79 &  97.07 \\
      8 &   6912 & 3.28739 & 338.03 & 163.63 \\
     12 &   2048 & 5.34723 & 292.53 & 100.60 \\
     16 &    864 & 5.53732 & 364.29 &  97.14 \\
    \bottomrule
  \end{tabular}
  \caption{Performance of the optimization level 8 kernel for
    different numbers of per-element grid points.  Note, the total
    number of grid points is equal between runs.  The \si{\flops} and
    bandwidth are estimated, not measured, as described in
    Section~\ref{sec:results}.
  }\label{tab:degree-dependencies}
\end{table}

\emph{Depdenency on \texttt{Nq}}.
Our next dataset highlights
the performance of the transformed kernel for
different values of \verb|Nq|, where the total number of degrees-of-freedom
is kept constant.  The results, given in
Table~\ref{tab:degree-dependencies}, show that the sensitivity of the
performance on \verb|Nq| varies depending on the hardware, with AMD hardware
tending towards greater sensitivity than Nvidia.
Ultimately, and irrespective of the target hardware, it is
quite likely that different tuning approaches may be needed
to yield consistent performacne even for each value of \verb|Nq|, even
on the same device--and transformation-based programming
with \loopy\ provides a clear path of deriving these
variants from the same, `clean' source code.

\medskip
The results presented in this section are for \loopy\footnote{The
kernels tested in Section~\ref{sec:results} are generated using
\loopy\ from the git repository
\url{http://git.tiker.net/trees/loopy.git} commit
\texttt{7d3f70aa76d627e95940c4281088dbc644da4cf5}.} generated
OpenCL kernels with 32-bit integers and floating point numbers.
The OpenCL build flags used with the AMD-APP (1912.5) and
CUDA 7.5.0 (355.11) platforms are:\\
\begin{tabular}{ll}
\verb|-cl-denorms-are-zero|, & \verb|-cl-fast-relaxed-math|, \\
\verb|-cl-finite-math-only|, & \verb|-cl-mad-enable|, and \\
\verb|-cl-no-signed-zeros|.
\end{tabular}\\
The performance measurements, \si{\flops} and bandwidth, are statically
counted in the generated kernels where add, multiply, divide, fused
multiply–add, and special functions count as 1 FLOP\@. For the
bandwidth calculation, all references (read or write) to global memory
are counted.  Due to caching effects, this way of measuring bandwidth
can exceed the peak global memory bandwidth rates for the given
devices. Timings were obtained using wall clock time by performing a few
untimed `warm-up' rounds of kernel execution, followed by the timing loop
which was run until an overall run time of at least 0.3 seconds was reached.

\acks%

AK's work on \loopy\ was supported in part by US Navy ONR grant number
N00014-14-1-0117, and by the National Science Foundation under grant
numbers DMS-1418961 and CCF-1524433. AK also gratefully acknowledges a
hardware gift from Nvidia Corporation. TW and LCW's work was supported
in part by US Navy ONR grant numbers N00014-15-WX-01603 and
N00014-13-1-0873, respectively.
The authors would like to thank M.~Wala for helpful suggestions that led
to improvements in the manuscript.

\clearpage
\FloatBarrier

\bibliographystyle{abbrvnat}


\bibliography{loopy,gnuma}

\end{document}